# T-junction ion trap array for two-dimensional ion shuttling, storage and manipulation

W. K. Hensinger, S. Olmschenk, D. Stick, D. Hucul,
M. Yeo, M. Acton, L. Deslauriers, and C. Monroe
*FOCUS Center and Department of Physics, University of Michigan, Ann Arbor MI, USA*

J. Rabchuk
*Department of Physics, Western Illinois University, Macomb IL, USA*

We demonstrate a two-dimensional 11-zone ion trap array, where individual laser-cooled atomic ions are stored, separated, shuttled, and swapped. The trap geometry consists of two linear rf ion trap sections that are joined at a 90 degree angle to form a T-shaped structure. We shuttle a single ion around the corners of the T-junction and swap the positions of two crystallized ions using voltage sequences designed to accommodate the nontrivial electrical potential near the junction. Full two-dimensional control of multiple ions demonstrated in this system may be crucial for the realization of scalable ion trap quantum computation and the implementation of quantum networks.

The ion trap stands out as a compelling quantum computing architecture[1], with significant recent progress in the implementation of quantum protocols using small numbers of trapped ion quantum bits (qubits)[2-6]. To scale the ion trap quantum computer beyond a few qubits, it may be necessary to operate arrays of ion traps where individual ions are shuttled between memory (storage) and interaction (entanglement) regions[1,7,8]. Shuttling of trapped ions along a line between adjacent trapping zones and separation of two or more ions has been demonstrated in a series of seminal experiments[9,3-5]. The next fundamental building block in this vision for an ion trap quantum computer is the reliable transport of ions through a multi-dimensional junction. In this paper we report the successful operation of a T-junction ion trap and demonstrate full 2-dimensional control of ion transport.

The main ion trap geometry used for quantum information processing is the linear rf-quadrupole trap[1], where ions are transversely confined to the nodal axis of an rf quadrupole potential supplied from nearby linear electrodes. Axial confinement can be accomplished by segmenting the linear electrodes and applying differential static potentials along the axis. In order to fabricate complex ion trap arrays, the electrodes can be fashioned from multi-layer planar substrates[7,10]. Asymmetric planar ion traps have also been proposed[11], where the electrodes do not surround the ions but lie in a plane nearby. In order to design a trapping geometry capable of supporting a two-dimensional junction, the electric field topology near the junction must be considered carefully. While two-layer geometries provide strong confinement in both transverse dimensions inside a linear chain of trapping regions, it is difficult to have sufficient transverse confinement in the junction region[11]. We use a symmetric three-layer geometry[6,7,10] (Fig. 1) that allows confinement throughout the junction region. The middle-layer carries the rf potential, and identical segmented outer-layers carry control (quasi-static) voltages that are used to confine the ion along the axial dimensions of the trap sections. Fig. 1 shows a top view and cross-section of the trap used in the experiments reported here.

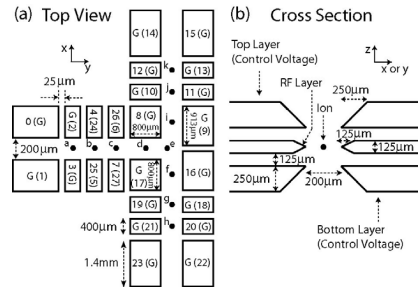

Fig.1. Top view and cross-section of two-dimensional trapping array. Dots depict the location of trapping zones a-k. The outer control electrodes are labeled, with bottom layer electrodes in parentheses. Electrodes labeled by "G" are internally grounded.

The T-junction trap is fabricated using thin, laser-machined polished alumina substrates similar to what has been used in previous one-dimensional linear traps[9,3-6,10] and is shown in Fig. 2. Ions can be confined in any of the 11 trapping zones labeled by letters *a-k* in Fig. 1. Ovens containing cadmium oxide are heated to produce a vapor of cadmium in the trapping region with an estimated partial pressure of under $10^{-11}$ torr. Cadmium ions are loaded by photoionizing the background vapor with a pulsed laser that is tuned near the atomic cadmium $^1S_0 \rightarrow {}^1P_1$ transition at 228.5 nm. Storage lifetimes of several hours are typical. The trapped ions are laser-cooled with a continuous-wave laser tuned near the $Cd^+$ $^2S_{1/2} \rightarrow {}^2P_{3/2}$ transition at 214.5 nm (radiative linewidth $\gamma/2\pi \approx 59$ MHz). The cooling laser is detuned several linewidths red of resonance with about 1 mW of power focused down to a ~15 μm waist. The ion is imaged with a CCD camera to a nearly diffraction-limited spot with f/2.1 optics. The imaging system was optimized to view an area of approximately 550 x 550 μm which allows for the simultaneous observation of trapping zones *d* and *i*, or *d* and *f* (Fig. 1), permitting real-time observation of a corner-turning shuttling protocol (Fig. 5). At this magnification, a diffraction-limited image of the point-like laser–cooled ion is a few pixels.

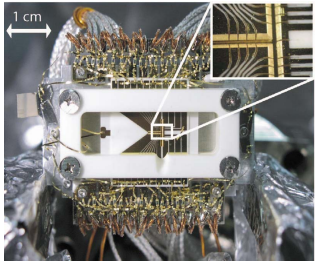

Fig. 2. T-junction trap array. The central layer contains a T-shaped channel; the electrode is formed by depositing gold around the channel with an electron beam evaporator. Gold-coating of the 24 control electrodes on each of the two outer layers is accomplished with dry-film photolithography and wet-chemical etching. Here, electrodes and tracks are formed by depositing 0.015 μm of titanium followed by 0.4 μm of gold. Two thin alumina spacer plates are inserted between each outer layer and the central rf layer substrate. All three substrates are held together via rectangular alumina mount bars. Chip capacitors and resistors are ribbon-bonded onto a gold coated quartz plate that is mounted adjacent to the alumina substrates (top and bottom of figure). To isolate the control electrodes from external noise and from induced rf from the nearby rf electrode, each of the 28 non-grounded control electrodes is immediately shunted to ground via a 1 nF capacitor and then connected in series to a 1 kΩ resistor leading to the vacuum feedthrough. The magnified inset shows the trapping array near the junction.

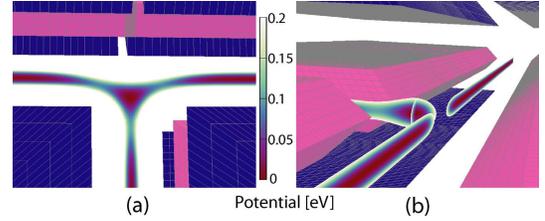

Fig. 3. Ponderomotive potential in the vicinity of the junction showing rf-nodes and three distinct humps. The contours are cutoff at 0.2 eV. The linear trap rf nodes leading into the junction (red) are clearly visible. Near the junction, there are small humps in the ponderomotive potential along any path leading to the junction (blue). The three humps have roughly equal heights (for the above applied rf voltage, hump in the stem region has $\Delta E \sim 0.1$ eV, humps in the top of the T have $\Delta E \sim 0.09$ eV) and each hump has a axial width (FWHM) of approximately 200 μm.

For the shuttling protocols reported here, an rf amplitude of $V_0 \approx 360$ V at $\Omega_T/2\pi = 48$ MHz is applied to the central layer electrode, resulting in a transverse ponderomotive secular frequency $\omega_{rf}/2\pi \sim 5.0$ MHz for the trapping zones *a, b, c, j, k, g,* and *h*. The more complicated ponderomotive potential near the junction is discussed below. The control voltages applied to the 28 control electrodes are computer-controlled[12]. Control voltages of order 100 V result in axial secular frequencies of order 2.5 MHz for traps whose central segment is 400 μm wide. Various composite trapped-ion shuttling protocols are implemented consisting of one or more key protocols: linear shuttling, corner turning, and separation/combination of two ions. Using these building blocks, high level shuttling procedures can be implemented.

The shuttling of ions through the T-junction merits special attention. Fig. 3 shows the calculated rf ponderomotive potential near the junction, providing transverse confinement throughout the shuttling path including the junction region. There are linear rf nodes leading towards the junction from all three directions that give way to small humps in the ponderomotive potential as the junction is approached, leading to a point node in the rf potential near the center of the T-junction (trapping zone *e* in Fig 1). These rf humps are small compared to the overall transverse ponderomotive potential walls, so time-varying voltages on the control electrodes can be used to push the ion over the rf humps. Shuttling a single atom around a corner requires a tradeoff: the time-varying pushing potentials must be strong enough to overcome the rf humps, but not so strong as to de-stabilize the trap in the transverse directions. The control voltage sequence must be carefully synchronized with the motion of the ion. Fast non-adiabatic voltage changes inside the trapping region may be required to minimize the kinetic energy acquired by the ion and to overcome the second hump upon emerging from the junction. Appropriate voltage sequences have been obtained by numerically solving Hamilton's equation[13]. Fig. 4 shows voltages applied to the electrodes carrying control voltages in order to shuttle the ion around the corner from trapping zone *d* to *i*. The pictorial sequence in Fig. 5 illustrates the resulting shuttling process. The success rate of the corner-turning protocol was measured to be essentially 100% (881 out of 882 attempts). Simulations predict that the ion acquires about 1.0 eV of kinetic energy during the corner-turning protocol. This energy is dissipated via Doppler cooling, but sympathetic cooling[1] can also remove this energy in order to preserve the internal state of the ion. It should be noted that, in principle, the gain in kinetic energy can be reversed with fast phase-sensitive switching of the trapping potentials without using any dissipative force[14].

In order to shuttle the ion back from the top of the T into the stem, a voltage sequence is used that corresponds approximately to the above corner-turning protocol but spatially reflected about the axis connecting electrodes 8 and 16. The success rate for this protocol was measured to be in excess of 98% (118 attempts). This sequence is conducted at slower speeds (20ms for the whole sequence) but refining the control voltage protocol may allow shuttling times on the order of microseconds[15]. Two other key protocols required for the implementation of universal shuttling operations are linear shuttling[3,4,5,9] and separation[9,16]. We performed linear shuttling operations with near-perfect efficiency over a wide range of speeds. A separation protocol is implemented inside the stem of the T array starting from zone b. Trapping zone b is weakened to $\omega_y/2\pi \sim 20$ kHz and a potential wedge separating the two ions is slowly brought up using electrodes 4 and 5, with electrodes 0, 1, 8 and 17 being used to confine the ion along the y-axis. Separation typically takes ~10ms and is carried out with a success rate of only ~58% (64 attempts), possibly limited by the very weak trap during separation and the large (400 μm) axial extent of the control electrodes[9]. Using these key protocols a composite protocol was successfully implemented for switching the position of two ions. The ions are separated in zone b, the first ion transferred to j, the second to h. The first ion is shuttled back to b. The second ion is shuttled back to b, having switched places with the first ion, with the two-ion chain effectively executing a "three-point turn." The protocol was carried out in multiple successive 10 ms steps. Conditional on successful separation and recombination, we obtain a success rate of 82% (34 attempts). The success rate for the whole process including separation and recombination is 24% (51 attempts), mainly limited by separation and recombination efficiency.[9]

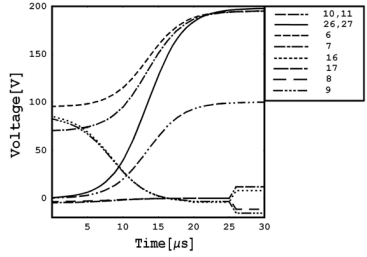

Fig.4. Control voltage schedule over time applied to the numbered electrodes to shuttle an ion around the corner from trap zone d to trapping zone i. The ion starts in trapping zone *d* at a voltage configuration corresponding to secular frequencies $\omega_x/2\pi \sim 5.0$ MHz, $\omega_y/2\pi \sim 0.7$ MHz and $\omega_z/2\pi \sim 4.9$ MHz. The first step is to shuttle the ion into the junction region *e*. This is achieved by simultaneously raising electrodes 6, 7, 26, 27 to 200V, lowering 9 and 16 to -2V and slightly raising 8 and 17 to 0V. Due to the shallow potential adjacent to the junction, the voltages are varied slowly (~20 μs) to stay in the adiabatic regime (change of secular frequency $d\omega/dt \ll \omega^2$). The final step consists of rapidly (~1 μs) raising electrodes 16 and 17 to 10V and simultaneously lowering electrodes 8 and 9 to –10V in order to shuttle the ion into trap *i* where it is trapped with secular frequencies $\omega_x/2\pi \sim 0.5$ MHz, $\omega_y/2\pi \sim 5.5$ MHz and $\omega_z/2\pi \sim 4.3$ MHz.

In conclusion, full two-dimensional control of the position of trapped atomic ions is demonstrated in a T-junction trap, including corner-turning and a protocol for swapping the positions of two ions. The two-dimensional control of multiple ions in this system could allow an efficient method for entangling arbitrarily-positioned (including non-adjacent) ions, with applications to large-scale trapped ion quantum computing. Future work will focus on characterizing the kinetic energy acquired during different shuttling operations and optimizing voltage sequences to minimize this kinetic energy and increase the reliability of the shuttling sequences. While the internal state of the ion is not expected to be significantly disturbed during the shuttling operation, future experiments will investigate qubit coherence in these protocols.

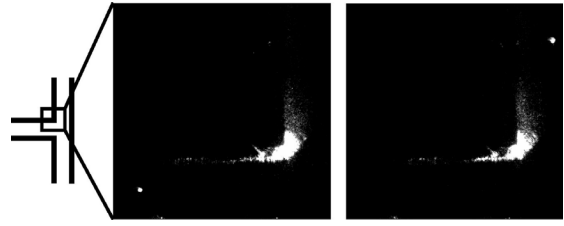

Fig. 5. Pictorial sequence illustrating how an ion is shuttled around the corner (from d to i).

We thank Rudolph N. Kohn Jr. and Jacob Burress for their valuable assistance. This work was supported by the U.S. National Security Agency and the Advanced Research and Development Activity under Army Research Office contract, and the National Science Foundation Information Technology Research Division.

3